\documentclass[aps,showpacs,twocolumn]{revtex4}%
\usepackage{amsfonts}
\usepackage{amsmath}
\usepackage{amssymb}
\usepackage{graphicx}%
\begin{document}
\title{Orbital magnetization and its effects in spin-chiral ferromagnetic kagom\'{e}
lattice in the general spin-coupling region}
\author{Zhigang Wang and Ping Zhang}
\affiliation{Institute of Applied Physics and Computational Mathematics, P.O. Box 8009,
Beijing 100088, P.R. China}
\pacs{75.30.-m, 73.43.-f, 72.15.Jf}

\begin{abstract}
The orbital magnetization and its effects on the two-dimensional kagom\'{e}
lattice with spin anisotropies included in the general Hund's coupling region
have been theoretically studied. The results show that the strength of the
Hund's coupling, as well as the spin chirality, contributes to the orbital
magnetization $\mathcal{M}$. Upon varying both these parameters, it is found
that the two parts of $\mathcal{M}$, i.e., the conventional part
$\mathbf{M}_{c}$ and the Berry-phase correction part $\mathbf{M}_{\Omega}$,
oppose each other. The anomalous Nernst conductivity is also calculated and a
peak-valley structure as a function of the electron Fermi energy is obtained.

\end{abstract}
\maketitle

\section{Introduction}

Recently the geometrically frustrated electron systems have provided hot
topics in the field of condensed matter physics \cite{Bramwell}. Ferromagnetic
pyrochlore $R_{2}$Mo$_{2}$O$_{7}$ ($R$=Nd, Sm, Gd) is one key type of the
geometrically frustrated systems \cite{Ramirez}, which consists of
corner-sharing tetrahedrons and the antiferromagnetic interactions between
nearest-neighbor spins are frustrated. It was recently pointed out that even
the ferromagnetic interaction is frustrated, if the spin easy axis points to
the center of the tetrahedron \cite{Harris}. In this case, the spin chirality
\cite{Kalmeyer}, which originated from the noncoplanar spin configuration, is
expected to affect the quantum of the electrons, especially the transverse
conductivity. This mechanism is also called \textquotedblleft Berry phase
contribution\textquotedblright\ because a non-vanishing spin chirality is
associated with a non-vanishing spin Berry phase for conduction electrons. To
interpret the transport experiments on ferromagnetic pyrochlore
\cite{Taguchi,Katsufuji}, Ohgushi et al. \cite{Ohgushi} studied the Hall
effect in a two-dimensional (2D) kagom\'{e} lattice, which is the cross
section of the porochlore lattice perpendicular to the (1,1,1) direction
\cite{Ramirez}. They obtained that if the chiral spin state is realized, the
system can show a quantized Hall effect. In their model an important limit is
used, which is that the electron conduction spins are colinear with lattice
(ions) spins. This limit is called that \textquotedblleft the strong (or
infinite) Hund's coupling limit\textquotedblright. However, detailed
experiments on the pyrochlores \cite{Taguchi2} show that the chiral mechanism
alone can not explain the anomalous transport phenomena in these systems. To
explain these experiments, Taillefumier et al. \cite{Taillefumier} studied the
same lattice in a general case, which extrapolates the strong and weak Hund's
coupling regions. They found that the spin Berry phase contribution does not
depend only on the spin chirality, but also on the strength of the local
Hund's coupling.

On the other hand, the orbital magnetism of Bloch electrons has been attracted
renewed interest, due to the recent recognition \cite{Xiao1,Thon1, Xiao2} that
the Berry phase effect plays an important role on orbital magnetism as well as
on the Hall conductivity. The Berry phase effect on orbital magnetism was
until now partially presented by very few studies
\cite{Xiao2,Lee,Thon2,Wang2007, Wang2}. Due to its basic importance in
understanding the magnetism and transport features of the materials,
obviously, more work are needed in exploiting the Berry phase effect on the
properties of the orbital magnetization (OM) in various kinds of realistic
physical systems.

In this paper we extend the study of the OM to the ferromagnetic pyrochlore
systems; more specially, we focus our attention to the 2D kagom\'{e} lattice
with spin anisotropies and Hund's coupling included. It is found that the two
parts in OM (see Sec. II), i.e., the conventional part $M_{c}$ and the
Berry-phase correction part $M_{\Omega}$, oppose each other. In particular,
the OM displays fully different behaviors in metallic and insulating regions
due to the different roles $M_{c}$ and $M_{\Omega}$ play in these two regions.
Moreover, similar to the role the Hund's coupling plays in determining the
anomalous Hall conductivity as observed in the above-mentioned experimental
\cite{Taguchi2} and theoretical \cite{Taillefumier} works, we find that the OM
is also importantly affected by the Hund's coupling. In particular, in the
weak coupling case that the Mott gap bewteen the upper and lower Hurbard bands
is overcome by the electron kinetic energy, we show that the OM exhibits
complex behaviors when scanning the Fermi energy through the whole series of
occupied bands. Furthermore, by using the obtained values of the OM we also
calculate the anomalous Nernst conductivity, which is featured by a
complicated peak-valley pattern as a function of the electron Fermi energy.

\section{Preliminaries}

Before studying the OM of the 2D kagom\'{e} lattice, we simply review the
general multiband formula for finite-temperature OM in the semiclassical
picture of Bloch electrons. In the semiclassical picture \cite{Chang, Sund},
the Bloch electron for the $n$th band is treated as a wave packet
$|w_{n}(\mathbf{r}_{c},\mathbf{k}_{c})\rangle$ with its center ($\mathbf{r}%
_{c},\mathbf{k}_{c}$) in the phase space. The orbital magnetic moment
characterizes the rotation of the wave packet around its centroid and is given
by $\mathbf{m}_{n}(\mathbf{k}_{c})$=$\frac{(-e)}{2}\langle w_{n}%
|(\mathbf{\hat{r}}-\mathbf{r}_{c})\times\mathbf{\hat{v}}|w_{n}\rangle$, where
$(-e)$ is the charge of the electron and $\mathbf{\hat{v}}$ is the velocity
operator. By writing the wave packet in terms of the Bloch state, one obtains
($\mathbf{k}_{c}$ is abbreviated as $\mathbf{k}$)%
\begin{equation}
\mathbf{m}_{n}(\mathbf{k})=-i(e/2\hbar)\langle\nabla_{\mathbf{k}%
}u_{n\mathbf{k}}|\times\lbrack\hat{H}_{\mathbf{k}}-\varepsilon_{n\mathbf{k}%
}^{(0)}]|\nabla_{\mathbf{k}}u_{n\mathbf{k}}\rangle, \label{e1}%
\end{equation}
where $|u_{n\mathbf{k}}\rangle$ is the periodic part of the Bloch state with
band energy $\varepsilon_{n\mathbf{k}}^{(0)}$, and $\hat{H}_{\mathbf{k}}$ is
the crystal Hamiltonian acting on $|u_{n\mathbf{k}}\rangle$. However, it was
further found \cite{Xiao1} that the presence of a weak magnetic field
$\mathbf{B}$ will result in a modification of the density of states in the
semiclassical phase space, $d^{3}\mathbf{k}\rightarrow d^{3}\mathbf{k}%
(1+e\mathbf{B}{\small \cdot}\mathbf{\Omega}_{n}/\hbar)$, where $\mathbf{\Omega
}_{n}(\mathbf{k})=i\langle\nabla_{\mathbf{k}}u_{n\mathbf{k}}|\times
|\nabla_{\mathbf{k}}u_{n\mathbf{k}}\rangle$ is the Berry curvature in
$k$-space. Due to this weak-field modification and the additional
thermodynamic average over Bloch bands included at finite temperature, the
total free energy for an equilibrium ensemble of electrons in the weak field
may be written as \cite{Xiao1}
\begin{equation}
F=-\frac{1}{\beta}\sum_{n}\int d^{3}\mathbf{k}\left(  1+\frac{e}{\hbar
}\mathbf{B}\cdot\mathbf{\Omega}_{n}(\mathbf{k})\right)  \ln[1+e^{\beta
(\mu-\varepsilon_{n\mathbf{k}})}]. \label{e2}%
\end{equation}
where $\mu$ is the electron chemical potential, $\beta=1/k_{B}T$ and
$\varepsilon_{n\mathbf{k}}$=$\varepsilon_{n\mathbf{k}}^{(0)}-\mathbf{m}%
_{n}(\mathbf{k})\cdot\mathbf{B}$ is the electron band energy in the presence
of the external magnetic field. The equilibrium OM density is given by the
field derivative at fixed temperature and chemical potential, $\mathcal{\vec
{M}}=-\left(  \partial F/\partial\mathbf{B}\right)  _{\mu,T}$, with the result%
\begin{align}
\mathcal{\vec{M}}  &  =\sum_{n}\int d^{3}\mathbf{km}_{n}(\mathbf{k}%
)f_{n}\nonumber\\
&  +\frac{1}{\beta}\sum_{n}\int d^{3}\mathbf{k}\frac{e}{\hbar}\mathbf{\Omega
}_{n}(\mathbf{k})\ln\left[  1+e^{\beta(\mu-\varepsilon_{n\mathbf{k}})}\right]
\nonumber\\
&  \equiv\mathbf{M}_{c}+\mathbf{M}_{\mathbf{\Omega}}, \label{e3}%
\end{align}
where $f_{n}$ is the local equilibrium Fermi function for $n$th band. In
addition to the conventional term $\mathbf{M}_{c}$ in terms of the orbital
magnetic moment $\mathbf{m}_{n}(\mathbf{k})$, the extra term $\mathbf{M}%
_{\mathbf{\Omega}}$ in Eq. (\ref{e3}) is a Berry phase effect and exposes a
new topological ingredient to the orbital magnetism. Interestingly, it is this
Berry phase correction that eventually enters the thermal transport current
\cite{Xiao2}. At zero temperature and magnetic field the general expression
(\ref{e3}) is reduced to \cite{Xiao1}%
\begin{equation}
\mathcal{\vec{M}}=\sum_{n}\int^{\mu_{0}}d^{3}\mathbf{k}\left(  \mathbf{m}%
_{n}(\mathbf{k})+\frac{e}{\hbar}\mathbf{\Omega}_{n}(\mathbf{k})\left[  \mu
_{0}-\varepsilon_{n\mathbf{k}}\right]  \right)  , \label{e4}%
\end{equation}
where the upper limit means that the integral is over states with energies
below the zero-temperature chemical potential (Fermi energy) $\mu_{0}$.

\section{Theoretical model and Chern number}

Following previous works \cite{Ohgushi,Taillefumier,Wang2007}, we consider the
double-exchange ferromagnet kagom\'{e} lattice schematically shown in Fig.
1(a). The triangle is the one face of the tetrahedron. Here we consider a pure
spin model with anisotropic Dzyaloshinskii-Moriya interactions on a kagom\'{e}
lattice. It consists of an umbrella of three spins per unit cell of the
kagom\'{e} lattice. Each umbrella can be described by the spherical
coordinates of the three spins ($\pi/6,\theta$), ($5\pi/6,\theta$), and
($-\pi/2,\theta$), as shown in Fig. 1(b). The angle $\theta$ ranges from $0$
to $\pi$.

The tight-binding model of this 2D kagom\'{e} lattice can be written as the
following \cite{Taillefumier}%
\begin{equation}
H=\sum_{\langle i,j\rangle,\sigma}t_{ij}\left(  c_{i\sigma}^{\dag}c_{j\sigma
}+\text{H.c.}\right)  -J_{0}\sum_{i,\alpha,\beta}c_{i\alpha}^{\dag}\left(
\mathbf{\sigma}_{\alpha\beta}\cdot\mathbf{n}_{i}\right)  c_{i\beta},
\label{Hamiltonian}%
\end{equation}
where $t_{ij}$ is the hopping integral between two neighboring sites $i$ and
$j$; $c_{i\sigma}^{\dag}$ and $c_{i\sigma}$ are the creation and annihilation
operators of an electron with spin $\sigma$ on the site $i$. $J_{0}$ is the
effective coupling constant to each local moment $\mathbf{S}_{i}$, and these
moments are treated below as classical variables. $\mathbf{n}_{i}$ is a unit
vector collinear with the local moment $\mathbf{S}_{i}$. $\mathbf{\sigma}$ are
the Pauli matrices. In the following we change notation $i\rightarrow(lms)$,
where $\left(  lm\right)  $ labels the kagom\'{e} unit cell and $s$ is the
site index in one unit cell. Note that in the infinite Hund's coupling limit,
i.e., $J_{0}\rightarrow\infty$, this system has already been discussed in
Refs. \cite{Ohgushi, Wang2007,Edge}. In this limit the two $\sigma$=$\uparrow
$, $\downarrow$ bands are infinitely split and the model describes a fully
polarized electron subject to a modulation of a fictitious magnetic flux.

To diagonalize the Hamiltonian (\ref{Hamiltonian}), we need to rewrite it in
the reciprocal space. We use the momentum representation of the electron
operator
\begin{equation}
c_{(lms\sigma)}=\frac{1}{\sqrt{L_{x}L_{y}}}\sum_{\mathbf{k}}e^{i\mathbf{k}%
\cdot\mathbf{R}_{(lms)}}\gamma_{s\sigma}(\mathbf{k}) \label{momentum}%
\end{equation}
and the one-particle state $|\Psi(\mathbf{k})\rangle$=$\sum_{s\sigma}$
$\Psi_{s\sigma}(\mathbf{k})\gamma_{s\sigma}^{\dag}(\mathbf{k})|0\rangle$.
Inserting $|\Psi(\mathbf{k})\rangle$ into the Schr\"{o}dinger equation
$H|\Psi\rangle$=$E|\Psi\rangle$, we can easily obtain the Hamiltonian in the
reciprocal space $H(\mathbf{k})$, which is given by\begin{widetext}
\begin{equation}
H(\mathbf{k})=\left(
\begin{array}
[c]{cccccc}%
-J_{0}\cos\theta & p_{\mathbf{k}}^{1} & p_{\mathbf{k}}^{3} & iJ_{0}\sin\theta
& 0 & 0\\
p_{\mathbf{k}}^{1} & -J_{0}\cos\theta & p_{\mathbf{k}}^{2} & 0 & -J_{0}%
\sin\theta e^{-i\frac{\pi}{6}} & 0\\
p_{\mathbf{k}}^{3} & p_{\mathbf{k}}^{2} & -J_{0}\cos\theta & 0 & 0 &
-J_{0}\sin\theta e^{-i\frac{5\pi}{6}}\\
-iJ_{0}\sin\theta & 0 & 0 & J_{0}\cos\theta & p_{\mathbf{k}}^{1} &
p_{\mathbf{k}}^{3}\\
0 & -J_{0}\sin\theta e^{i\frac{\pi}{6}} & 0 & p_{\mathbf{k}}^{1} & J_{0}%
\cos\theta & p_{\mathbf{k}}^{2}\\
0 & 0 & -J_{0}\sin\theta e^{i\frac{5\pi}{6}} & p_{\mathbf{k}}^{3} &
p_{\mathbf{k}}^{2} & J_{0}\cos\theta
\end{array}
\right) ,\label{Hk}
\end{equation}
\end{widetext}where $t$=$|t_{ij}|$ as the energy unit and $p_{\mathbf{k}}^{i}%
$=$2t\cos\left(  \mathbf{k\cdot a}_{i}\right)  $. $\mathbf{a}_{1}$=$(-\frac
{1}{2},-\frac{\sqrt{3}}{2})$, $\mathbf{a}_{2}$=$(1,0)$, and $\mathbf{a}_{3}%
$=$(-\frac{1}{2},\frac{\sqrt{3}}{2})$ represent the displacements in a unit
cell from A to B site, from B to C site, and from C to A site respectively. In
this notation, the Brillouin zone (BZ) is a hexagon with the corners of
$\mathbf{k}=\pm(2\pi/3)\mathbf{a}_{1}$, $\pm(2\pi/3)\mathbf{a}_{2}$, $\pm
(2\pi/3)\mathbf{a}_{3}$, two of which are independent.

\begin{figure}[ptb]
\begin{center}
\includegraphics[width=1.0\linewidth]{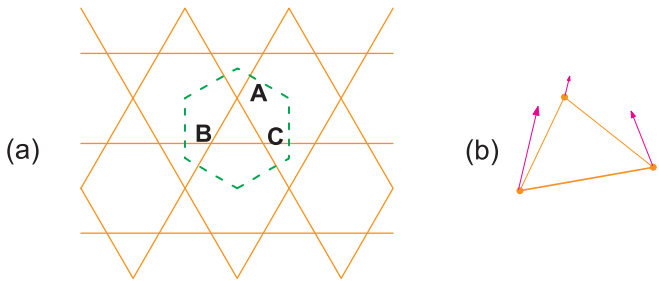}
\end{center}
\caption{(Color online) (a) Two dimensional spin-chiral ferromagnetic
kagom\'{e} lattice. The dashed line represents the Wigner-Seitz unit cell,
which contains three independent sites (A, B, C). (b) The umbrella structure
on the triangular cell of the 2D kagom\'{e} lattice. }%
\end{figure}\begin{figure}[ptbptb]
\begin{center}
\includegraphics[width=1.0\linewidth]{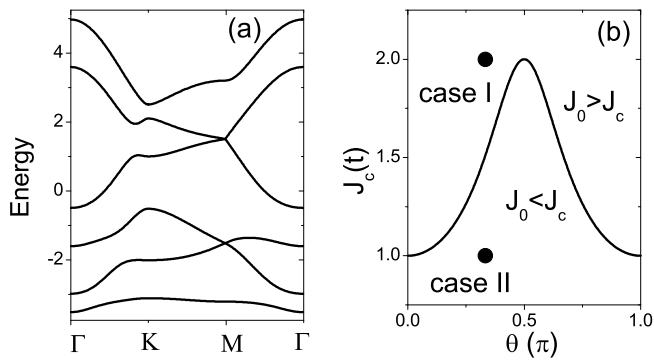}
\end{center}
\caption{(a) Energy spectrum of the 2D kagom\'{e} lattice with the parameters
as $\theta$=$\pi/3$, $J_{0}$=$J_{c}$=$4/\sqrt{7}$. (b) The critical value of
Hund's coupling $J_{c}$ as a function of the chiral parameter $\theta$.}%
\end{figure}

Now let us consider the Chern number \cite{Thouless} and Hall conductivity of
this system, which have been reported in Refs. \cite{Taillefumier, Edge2}. In
the strong Hund's coupling limit, there is an energy gap between the two
nearest-neighbor bands in the cases of $\theta\neq0$, $\pi$. Then the Hall
conductivity is a sum over occupied Bloch bands,
\begin{equation}
\sigma_{xy}=\frac{e^{2}}{h}\sum_{n}^{\text{occu}.}C_{n}, \label{1}%
\end{equation}
where the $n$th band Chern number is defined by%
\begin{equation}
C_{n}=-\frac{1}{2\pi}\int_{\text{BZ}}d^{2}\mathbf{k}\Omega_{n}\left(
\mathbf{k}\right)  =-\frac{1}{2\pi}\int_{\text{BZ}}d^{2}\mathbf{k}\hat{z}%
\cdot\left[  \nabla_{\mathbf{k}}\times\mathbf{A}_{n}\left(  \mathbf{k}\right)
\right]  , \label{Chern}%
\end{equation}
where $\mathbf{A}_{n}\left(  \mathbf{k}\right)  $=$i\langle u_{n\mathbf{k}%
}|\nabla_{\mathbf{k}}u_{n\mathbf{k}}\rangle$ is the Berry-phase connection
(vector potential) for the $n$th band. At finite temperature, considering the
electron density distribution, the Hall conductivity is written as
\begin{equation}
\sigma_{xy}=-\frac{e^{2}}{h}\int_{\text{BZ}}\frac{d^{2}\mathbf{k}}{2\pi}%
f_{n}\hat{z}\cdot\left[  \nabla_{\mathbf{k}}\times\mathbf{A}_{n}\left(
\mathbf{k}\right)  \right]  . \label{Hallconductivity}%
\end{equation}
However in the general spin coupling cases, the gap between two
nearest-neighbor bands may disappear. When the Fermi energy lies in these two
bands, becasue the gap vanishes, the Hall conductivity can not be written in
the form of Eq. (\ref{1}). In despite of this, the concept of the $n$th band
Chern number and Eq. (\ref{1}) are also useful when the gap between two
nearest-neighbor bands does not disappear.

For the general cases the energy spectrum can only be computed numerically,
except for general $\theta$ at high-symmetry points. For the finite values of
$J_{0}$, as pointed in \cite{Taillefumier}, the splitting of the spectrum
depends on two mechanisms. One is that the coupling $J_{0}$ separates each
group of three bands, the other is that when switching on $J_{0}$, the
pointlike degeneracies are lifted within each group of three bands. According
to the properties of the $M$ point of the Brillouin zone, a critical value of
the Hund's coupling as a function of the chiral parameter $\theta$ can be
analytically obtained, which is given by \cite{Taillefumier},%
\begin{equation}
J_{c}(\theta)=\pm\frac{2t}{\sqrt{1+3\cos^{2}\theta}}. \label{critical}%
\end{equation}
Using Eq. (\ref{critical}) one can distinguish between two different regimes
depending on the value of $J_{0}$ as compared to $J_{c}(\theta)$. In the
regime where $J_{0}>J_{c}$, the Chern numbers associated with each band are
given by $-1$, $0$, $1$, $1$, $0$, $-1$ from the lowest to the topmost band.
Whereas in the regime where $J_{0}<J_{c}$, the Chern numbers associated with
each band are given by $-1$, $3$, $-2$, $-2$, $3$, $-1$ in the same order. We
draw in Fig. 2(a) the energy spectrum with the spin chiral parameter $\theta
$=$\pi/3$ and $J_{0}$ taking the critical value $J_{c}$=$4/\sqrt{7}$. Fig.
2(b) shows the critical value of the Hund's coupling $J_{c}$ as a function of
the chirality $\theta$.

\section{The orbital magnetization}

Now we turn to study the OM of the 2D kagom\'{e} lattice in the
general Hund's coupling cases. Similar to the Hall conductivity, the
OM displays different behaviors in two regions which we will exhibit
in turn. As examples, we consider two cases. The case I, in which we
set the parameters as $\theta$=$\pi/3$ and $J_{0}$=$2$, is one
typical case in the regime where
$J_{0}>J_{c}$(=$4/\sqrt{7}\approx1.51$). Whereas the case II, in
which the parameters are $\theta$=$\pi/3$ and $J_{0}$=$1$, is
another typical case in the regime where $J_{0}<J_{c}$. These two
cases are shown in Fig. 2(b) with solid dots.

First we consider the case I. To more clearly investigate the OM of the 2D
kagom\'{e} lattice, we need to know the energy band structure of the system.
So, we draw the energy spectrum in Fig. 3(a), from which one can find that
there are four gaps. From the lowest to the topmost gap, we denote these gaps
as gap-I, -II, -III, and -IV. Clearly, in this case only the gap between bands
$5$ and $6$ vanishes. Fig. 3(b) plots the Hall conductivity as a function of
the electron Fermi energy (chemical potential). \begin{figure}[ptb]
\begin{center}
\includegraphics[width=1.0\linewidth]{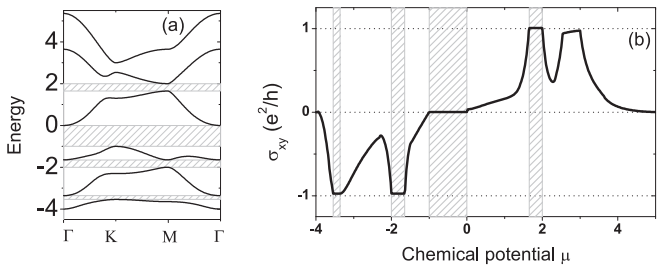}
\end{center}
\caption{(a) The energy spectrum of the 2D kagom\'{e} lattice. (b) The Hall
conductivity $\sigma_{xy}$ as a function of the chemical potential $\mu$. In
both figures, the chiral parameter is $\theta$=$\pi/3$ and the strength of the
Hund's coupling $J_{0}$=$2$. The shaded areas are the energy gaps, which
labeled as gap-I, -II, -III, and -IV from the lowest to the topmost gap.}%
\end{figure}\begin{figure}[ptbptb]
\begin{center}
\includegraphics[width=0.8\linewidth]{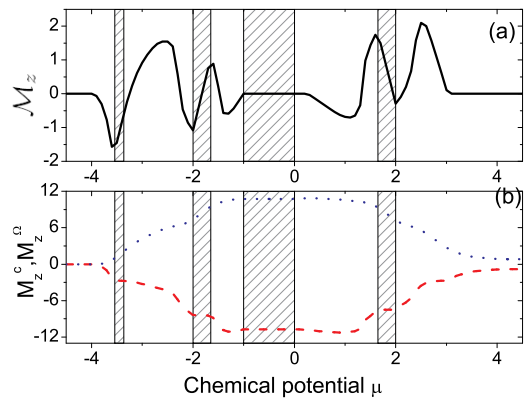}
\end{center}
\caption{(Color online) (a) The OM and (b) its two components $M_{c}$ (dashed
line) and $M_{\Omega}$ (dotted line) as a function of the chemical potential
$\mu$. The parameters are same as those in Fig. 3.}%
\end{figure}

Figure 4(a) shows the OM ($\mathcal{M}$)\ as a function of the electron
chemical potential $\mu$. One can see that initially the OM rapidly decreases
as the filling of the lowest band increases, arriving at a minimum at $\mu
$=$-3.54$, a value corresponding to the top of the lowest band. Then, as the
chemical potential continues to vary in the gap-I, the OM goes up and
increases as a linear function of $\mu$. This linear relationship in the
insulating region can be understood by Eq. (\ref{e3}), from which one obtains
\begin{align}
\frac{d\mathcal{M}}{d\mu}  &  =\frac{e}{\hbar}\sum_{n}^{\text{occu}}\int
d^{2}\mathbf{k}\Omega_{n}(\mathbf{k})\label{om-chem}\\
&  =-\frac{e}{h}\sum_{n}^{\text{occu}}C_{n}.\nonumber
\end{align}
Thus when the chemical potential varies in the gap-I, only the lowest band is
occupied and $d\mathcal{M}/d\mu=-(e/h)C_{1}$. In this case, $C_{1}=-1$. Thus
$d\mathcal{M}/d\mu=e/h$, i.e., the OM linearly increases with the chemical
potential in the insulating region I, as shown in Fig. 4(a). Similarly, when
the chemical potential increases in the gap-II, the OM increases linearly with
$\mu$. Since the Chern number of band $2$ is zero, thus from Eq.
(\ref{om-chem}) and Fig. 4(a) one can see that the slope of the OM curve in
the gap-II is same as that in the gap-I. When the chemical potential increases
in the gap-III, the OM becomes zero and does not varies with the chemical
potential. The reason is that the sum over the Chern numbers of the occupied
lowest three bands is zero. From Eq. (\ref{om-chem}), one can see that the
slope of the OM is independent of $\mu$. In fact the gap-III is the usually
called the Mott gap. With the chemical potential increases, when it lies in
the gap-IV, one can find that the OM decreases linearly with $\mu$. The reason
is that the sum of the Chern numbers of the lowest four bands is $1$. From Eq.
(\ref{om-chem}), one can find $d\mathcal{M}/d\mu=-e/h$, i.e., the OM linearly
decreases with the chemical potential in the gap-IV as shown in Fig. 3(b).

For further study, we show in Fig. 4(b) $M_{c}$ and $M_{\Omega}$ as a function
of the chemical potential, their sum gives $\mathcal{M}$ in Fig. 4(a). One can
see that overall $M_{c}$ and $M_{\Omega}$ have opposite contributions to
$\mathcal{M}$, which implies that these two parts carry opposite-circulating
currents. In each insulating area the conventional term $M_{c}$ keeps a
constant, which is due to the fact that the upper limit of the $k$-integral of
$m_{n}(\mathbf{k})$ is invariant as the chemical potential varies in the gap.
In the metallic region, however, since the occupied states varies with the
chemical potential, thus $M_{c}$ also varies with $\mu$, resulting in a
decreasing slope in the lowest three metallic regions and a increasing slope
in the highest two metallic regions, as shown in Fig. 4(b). The Berry phase
term $M_{\Omega}$ also displays different behavior between insulating and
metallic regions. In the lowest two insulating region, $M_{\Omega}$ linearly
increases with $\mu$, and in the gap-IV, $M_{\Omega}$ linearly decreases with
$\mu$, as is expected from Eq. (\ref{e3}). In the metallic region, however,
this term sensitively depends on the topological property of the band in which
the chemical potential is located. For the bands with nonzero Chern number,
one can see from Fig. 4(b) that $M_{\Omega}$ remains invariant, while for the
band $2$ (its Chern number is zero), it increases with the chemical potential
$\mu$. On the whole the comparison between Figs. 4(a) and 4(b) shows that the
metallic behavior of $\mathcal{M}$ is dominated by its conventional term
$M_{c}$, while in the insulating regime $M_{\Omega}$ plays a main role in
determining the behavior of $\mathcal{M}$. The behavior of the OM in this case
is similar to that in the strong spin coupling limit.

\begin{figure}[ptb]
\begin{center}
\includegraphics[width=1.0\linewidth]{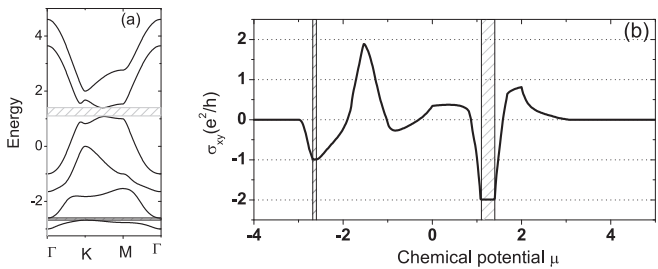}
\end{center}
\caption{(a) The energy spectrum of the 2D kagom\'{e} lattice. (b) The Hall
conductivity $\sigma_{xy}$ as a function of the chemical potential $\mu$. In
both figures, the chiral parameter is $\theta$=$\pi/3$ and the strength of the
Hund's coupling $J_{0}$=$1$. The shaded areas are the energy gaps, which
labeled as the lower band and the higher band, respectively.}%
\end{figure}

Then we study the case II. Similar to the case I, we firstly draw the energy
spectrum in Fig. 5(a), from which one can find that there are only two gaps,
which can be called the lower and the higher gap, respectively. In this case
the gaps between bands $2$ and $3$, between $3$ and $4$ (the Mott gap), and
between $5$ and $6$ vanish. We also draw in Fig. 5(b) the corresponding Hall
conductivity as a function of the electron Fermi energy. \begin{figure}[ptb]
\begin{center}
\includegraphics[width=0.8\linewidth]{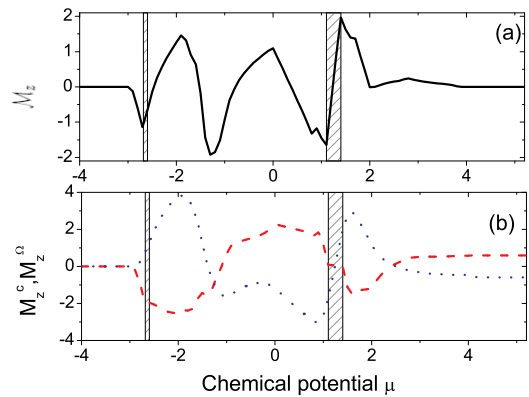}
\end{center}
\caption{(Color online) (a) The OM and (b) its two components $M_{c}$ (dashed
line) and $M_{\Omega}$ (dotted line) as a function of the chemical potential
$\mu$. The parameters are same as those in Fig. 5.}%
\end{figure}

Figure 6(a) plots the OM ($\mathcal{M}$)\ as a function of the electron
chemical potential $\mu$. One can see that initially the OM rapidly decreases
as the filling of the lower band increases, arriving at a minimum at $\mu
$=$-2.68$, a value corresponding to the top of the lower band. Then, as the
chemical potential continues to vary in the lower gap, the OM goes up and
increases as a linear function of $\mu$. This linear relationship in the
insulating region can also be understood by Eq. (\ref{om-chem}). Because the
gap between $2$ and $3$ and the Mott gap vanish, the OM displays a more
complex behavior in this metallic region. The value of the OM oscillate versus
the chemical potential $\mu$. When $\mu$ reaches the bottom of the higher gap,
the OM begins to go up again and linearly increases. Different from in the
lower gap, the coefficient of the increasing is twice of that in the lower
gap, which can be understood by Eq. (\ref{om-chem}). In the higher gap
$d\mathcal{M}/d\mu$=$-(e/h)\sum_{n=1}^{4}C_{n}$=$2e/h$, while in the lower
gap, $d\mathcal{M}/d\mu$=$-(e/h)C_{1}$=$e/h$.

To see the different roles $M_{c}$ and $M_{\Omega}$ play in the
metallic and insulating regions, we draw in Fig. 6(b) $M_{c}$ and
$M_{\Omega}$ as a function of the chemical potential, their sum
gives $\mathcal{M}$ in Fig. 6(a). In each insulating area the
conventional term $M_{c}$ keeps a constant and $M_{\Omega}$ linearly
increases with $\mu$ in the two gaps with different linear
coefficients, as is expected from Eq. (\ref{e3}). The behaviors of
OM in this case is novel, comparing with those in the infinite
Hund's coupling case. There are two features in the regime
$J_{0}<J_{c}$. One is that the magnitudes of both two parts $M_{c}$
and $M_{\Omega}$ become much smaller than those in the regime
$J_{0}>J_{c}$. The other is that in the metallic region, both two
parts rapidly changes. The reason is that the gaps between bands 2,
3 and 4 vanish and these bands form one energy band. However from
Fig. 6(b) one can see that the two parts of $\mathcal{M\ }$have
opposite contributions to $\mathcal{M}$, which implies that these
two parts carry opposite-circulating currents.

\section{Anomalous Nernst Effect}

The above discussion of $M_{c}$ and $M_{\Omega}$ can be transferred to study
the ANE. The relation between the OM and ANE has been recently found
\cite{Xiao2}. To discuss the transport measurement, it is important to
discount the contribution from the magnetization current, a point which has
attracted much discussion in the past. Motivated by the argument \cite{Cooper}
that the magnetization current cannot be measured by conventional transport
experiments, Xiao et al. \cite{Xiao2} have built up a remarkable picture that
the conventional orbital magnetic moment $M_{c}$ does not contribute to the
transport current, while the Berry phase term in Eq. (\ref{e2}) directly
enters and therefore modifies the intrinsic transport Hall current equation as
follows%
\begin{equation}
\mathbf{j}_{\text{H}}\mathbf{=}-\frac{e^{2}}{\hbar}\mathbf{E}\times\sum
_{n}\int\frac{d^{2}k}{(2\pi)^{2}}f_{n}(\mathbf{r},\mathbf{k})\Omega
_{n}(\mathbf{k})\mathbf{-}\nabla\times\mathbf{M}_{\Omega}(\mathbf{r}),
\label{jh}%
\end{equation}
In the case of uniform temperature and chemical potential, obviously, the
second term is zero and the Hall effect of 2D kagom\'{e} lattice is featured
by nonzero Chern number as discussed by Taillefumier et al.
\cite{Taillefumier} as well as in this paper. In the following, however, we
turn to study another situation, where the current-driving force is not
provided by the electric field ($\mathbf{E}$=0). Instead, it is provided by a
statistical force, i.e., the gradient of temperature $T$. In this case, Eqs.
(\ref{jh}) and (\ref{e2}) give the expression of intrinsic thermoelectric Hall
current as $j_{x}=\alpha_{xy}(-\nabla_{y}T)$, where the anomalous Nernst
conductivity $\alpha_{xy}$ is given by \cite{Xiao2}%
\begin{align}
\alpha_{xy}  &  =\frac{1}{T}\frac{e}{\hbar}\sum_{n}\int\frac{d^{2}k}%
{(2\pi)^{2}}\Omega_{n}\label{ane}\\
&  \times\left[  \left(  \epsilon_{n\mathbf{k}}-\mu\right)  f_{n}+k_{B}%
T\ln\left(  1+e^{-\beta(\epsilon_{n\mathbf{k}}-\mu)}\right)  \right]
.\nonumber
\end{align}
\begin{figure}[ptb]
\begin{center}
\includegraphics[width=0.6\linewidth]{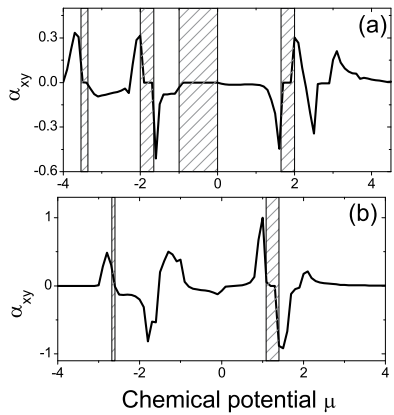}
\end{center}
\caption{The ANE of the 2D kagom\'{e} lattice at $k_{B}T$=$0.005$. The
parameters are (a) $\theta$=$\pi/3$ and $J_{0}$=$2$; (b) $\theta$=$\pi/3$ and
$J_{0}$=$1$. The shaded areas are the energy gaps.}%
\end{figure}

Figure 7 shows $\alpha_{xy}$ of the 2D kagom\'{e} lattice as a function of the
chemical potential for $k_{B}T$=$0.005$. One can see that the ANE disappears
in the insulating regions, and when scanning the chemical potential through
the bands, there will appear peaks and valleys. Remarkably, a similar
peak-valley structure was also found by the recent first-principles
calculations in CuCr$_{2}$Se$_{4-x}$Br$_{x}$ compound \cite{Xiao2}. The ANE of
this compound was recently measured by Lee et al. \cite{Lee} as a function of
Br doping $x$ which is used to change the chemical potential $\mu$. Due to the
scarce data available, until now the peak-valley structure of $\alpha_{xy}$
revealed in Fig. 7 and in Ref. \cite{Xiao2,Wang2007, Wang2} has not been found
in experiment, and more direct experimental results are needed for
quantitative comparison with the theoretical results. Interestingly, the
expression for $\alpha_{xy}$ can be simplified at low temperature as the Mott
relation \cite{Xiao2},
\begin{equation}
\alpha_{xy}=-\frac{\pi^{2}}{3}\frac{k_{B}^{2}T}{e}\frac{\partial\sigma
_{xy}(\mu_{0})}{\partial\mu_{0}}. \label{Mott}%
\end{equation}
Thus one can see that unlike the anomalous Hall effect \cite{Haldane}, ANE is
given by the Fermi-surface contribution of the band structure and Berry
curvature. Another unique feature of $\alpha_{xy}$ is its linear dependence of temperature.

\section{Summary}

We have theoretically studied the OM and ANE of the 2D kagom\'{e} lattice with
spin anisotropies included in a general Hund's coupling region, as a
supplement of the previous work \cite{Wang2007} in the infinite Hund's
coupling limit. The results show that both of the strength of the Hund's
coupling and the chirality contribute to the orbital magnetization
$\mathcal{M}$. Upon varying both these parameters, it is found that the two
parts of $\mathcal{M}$, i.e., the conventional part $\mathbf{M}_{c}$ and the
Berry-phase correction part $\mathbf{M}_{\Omega}$, oppose each other. We also
calculate the anomalous Nernst conductivity and obtain a peak-valley structure
as a function of the electron Fermi energy.

\begin{acknowledgments}
This work was supported by NSFC under Grants Nos. 10604010 and 10544004.
\end{acknowledgments}

\end{document}